\documentclass[a4paper]{jpconf}
\usepackage{graphicx}

\usepackage{amssymb,amsmath}
\usepackage{epsfig}
\usepackage{color}

\begin{document}
\title{Primordial power spectrum features and consequences}

\author{Gaurav Goswami}

\address{IUCAA, Post Bag 4, Ganeshkhind, Pune-411007, India}

\ead{gaurav@iucaa.ernet.in}

\begin{abstract}
The present Cosmic Microwave Background (CMB) temperature and polarization anisotropy data is consistent with 
not only a power law scalar primordial power spectrum (PPS) with a small running but also with the scalar PPS
having very sharp features. This has motivated inflationary models with such sharp features. Recently, even the 
possibility of having nulls in the power spectrum (at certain scales) has been considered.
The existence of these nulls has been shown in linear perturbation theory.
What shall be the effect of higher order corrections on such nulls? Inspired by this question, 
we attempt to calculate quantum radiative corrections to the Fourier transform of the two-point function in a toy field theory
and address the issue of how these corrections to the power spectrum behave in models in which
the tree-level power spectrum has a sharp dip (but not a null). In particular, we consider the possibility of the relative enhancement of
radiative corrections in a model in which the tree-level spectrum goes through a dip in power at a certain scale.
The mode functions of the field (whose power spectrum is to be evaluated) are chosen such that they undergo 
the kind of dynamics that leads to a sharp dip in the tree level power spectrum. Next we consider the situation 
in which this field has quartic self interactions and find one loop correction in a suitably chosen
renormalization scheme. 
We thus attempt to answer the following key question in the context of this toy model (which is as important in
the realistic case): in the chosen renormalization scheme, can quantum radiative corrections 
get enhanced relative to tree-level power spectrum at scales at which sharp dips appear in the tree-level spectrum?
\end{abstract}

It is a well known fact that the correlations of CMB anisotropies depend not only on the values of various (late time) 
background cosmological parameters (e.g. $\Omega_b$, $\Omega_c$ etc) but also on the assumed form of 
the PPS. 
Cosmological inflation 
has been a very 
actively investigated paradigm for explaining the origin of primordial metric perturbations that lead to 
anisotropies in CMB sky as well as to the large scale structure of the universe. There exist models of inflation
that give a smooth, nearly scale invariant scalar PPS (with possibly small running of the spectral index). But 
there also exist models 
\cite{PI1, Staro1992} 
in which the PPS is not so smooth, so has very sharp features. Theoretically, 
at this stage, there is no way to favour models that give smooth (featureless) PPS with those that do not. Interestingly, at this 
stage, even observationally, it is not possible to favour one of these. 
In many such theoretical scenarios, the scalar PPS has cuspy dips~
that sometimes correspond to a null in the PPS i.e. precisely zero scalar power at some wave number. 
Also for a range of modes near such a feature, the tensor power overtakes scalar
power.  
Such cusp-like dips in scalar PPS were reported in the literature but their origin
was not satisfactorily understood until recently (see \cite{GG-TS-2010} and references therein). 
An exact null in scalar PPS can have interesting consequences, such as on processed non-linear matter power spectrum. 
This leads to an interesting possibility: since the power spectrum calculation is done perturbatively, and since the leading order answer 
is too small (or zero) at some scale, could the higher order corrections be important at this scale? If there is no power at some scale, 
can higher order corrections become so important that they become dominant at this scale? 
We attempt to answer such questions in this paper in the context of a toy field theory.

Recall that in cosmological perturbation theory one is concerned with the theory of fluctuations around a background (inflationary) 
solution. One quantizes these fluctuations and calculates their correlation functions (on a constant time hyper surface) using the 
(well known) in-in formalism. The leading higher order corrections to the two-point function involve the one loop corrections. 
In the most extreme case of our interest, the tree-level contribution to the two-point function is zero and the 
power spectrum at this scale gets generated purely by radiative corrections. We are interested in the two-point function,
\begin{equation} \label{G}
  \langle \phi(\eta, \vec{x}_1) \phi (\eta, \vec{x}_2) \rangle = \int \frac{d^3 \vec{k}_1}{(2\pi)^3} \frac{d^3 \vec{k}_2}{(2\pi)^3} 
e^{i(\vec{k}_1\cdot \vec{x}_1 + \vec{k}_2\cdot \vec{x}_2)} \times (2\pi)^3 \delta^3(\vec{k}_1+\vec{k}_2) {\cal G} (k_1,\eta)
\end{equation}

\noindent (where the delta function on the RHS is due to spatial homogeneity of the background).
The connected part of the quantity ${\cal G}$, introduced in Eq. (\ref{G}), can be evaluated perturbatively
\begin{equation}
 {\cal{G}}(p,\eta) = C_0(p,\eta) + C_1(p,\eta) \lambda + C_2(p,\eta) {\lambda}^2 + \cdots
\end{equation}

\noindent Since the limit $\lambda \rightarrow 0$ recovers the free theory, $C_0(p,\eta) = |\phi_p(\eta)|^2$ where 
$\phi_p(\eta)$ is the mode function which determines the tree level PPS. 
In this article, we will attempt to calculate $C_1(p,\eta)$ perturbatively.

%
%

To illustrate the effect we are interested in,
we will try to calculate correlation functions in a field theory which mimics the dynamics 
of comoving curvature perturbation ${\cal R}(\eta,\vec{x})$. We solve for the correlations of a fictitous field $\sigma$ with 
the following (classical) action:
\begin{equation} \label{sigaction}
 S = \int d\eta~ d^3 {\vec{x}}~ z^2 \left( \frac{{(\sigma')}^2}{2} - \frac{(\vec{\nabla} {\sigma})^{2}}{2} 
- z^2 \frac{\lambda {\sigma}^4}{4!} \right)
\end{equation}

\noindent where we have replaced all occurances of $a$ in the action of a test scalar field (on an expanding unvierse) by 
$z= - a \sqrt{2 \epsilon}$ (here $a$ is the scale factor and $\epsilon$ is the slow-roll parameter)
and the function $z(\eta)$ shall be chosen such that the tree-level power spectrum has sharp features.
The realistic case is quite difficult to deal with, so we have made a few simplifying assumptions. 
To begin with, let us assume quartic self interactions of the fluctuation field (whose power spectrum (PS) has a 
sharp dip at tree level). This is because the one-loop one-vertex diagram will go as ${\cal R}^4$ while the 
one-loop two-vertex diagram will go as ${\cal R}^6$, ${\cal R}$ being the comoving curvature perturbation. 
It is important to realize that the action in Eq. (\ref{sigaction}) is not as arbitrary as it may look naively. When one writes down the 
fourth-order action for fluctuations (e.g. comoving curvature perturbation), one gets way too many terms 
(see e.g. \cite{S4wosr} where this is done without assuming slow-roll 
approximation for $\delta \phi$ in spatially flat gauge, in this gauge ${\cal R} = \frac{H}{\dot{\phi}} \delta \phi$, so we can express 
the action in terms of $\cal R$). The term with quartic self-interactions is one among many of these terms (corresponding to the first 
term in Eq. (36) of \cite{S4wosr}, notice that the coefficient of quartic term is such that we will get the factor of $z^4$ in the 
coefficient as is assumed above), and we shall be finding out what will happen to the loop corrections to the power spectrum due to 
this term alone. 
As a second simplifying assumption, let us attempt the case in which though the 
tree level PS has a sharp dip, this dip does not lead to a null in the PS. This will ensure that the one-loop 
correction is not dominant over the tree level answer, only gets enhanced (thus perturbation theory will still be valid).
The quadratic part of the above action determines the tree level power spectrum and 
we choose the dynamics of $z$ such that the tree level power spectrum has a sharp dip, but not a null (we choose $z$ to take the 
form in takes in Starobinsky model \cite{Staro1992}).

One can canonically quantize such a theory in the usual way. We will be calculating the correlation functions in canonical 
formalism in an interaction picture in which the time-dependence of the field operators shall be governed by the part of the Hamiltonian 
quadratic in the the field and the field satisfies a linear differential equation. Given the mode functions of these free fields, 
any correlation function can be evaluated.

While evaluating correlation functions at one loop, one encounters (just like in flat spacetime field theory) divergences.
Apart from the usual UV divergences and IR divergences (the comoving curvature perturbation behaves very much like 
a massless test field), one also encounters (extra) late time divergences. This is because when we evaluate the 
two-point function in the in-in formalism (for an interaction with an even power of field) we get

\begin{equation} 
\begin{split}
  \langle \Omega | \phi_H(\eta, \vec{x}_1) \phi_H (\eta, \vec{x}_2) |\Omega \rangle =  
\langle 0 | \phi_I(\eta, \vec{x}_1) \phi_I (\eta, \vec{x}_2) |0 \rangle \\
+ i  \int_{-{\infty}}^{\eta} d\eta' a(\eta')  
\langle 0 | [H_I(\eta'),\phi_I(\eta, \vec{x}_1) \phi_I (\eta, \vec{x}_2),] |0 \rangle + \cdots
\end{split}
\end{equation}

\noindent and the contribution when the external time $\eta$ is sent to zero diverges.

Due to the presence of the above stated late time divergences, it is natural that the loop corrections to the correlations 
cannot be kept small. There is a vast literatue on this subject, attempting to understand the origin and resolution of these 
divergences. It is important to understand that this is not the kind of enhancement we are interested in: 
we are talking about a situation in which the loop correction gets enhanced relative 
to tree level result because the tree level contribution itself has a sharp dip with very low power. To get rid of the late time 
divergences, we shall cut the (one-loop) time integral. The prescription (which we use) to do this is that just like the tree 
level PS, even the loop correction to the power spectrum for the field on de-Sitter background is scale invariant.

It is important to note that if one takes into account all the self interactions of the metric fluctuations (e.g. the comoving 
curvature perturbation) dictated by the symmetries of the problem and evaluates 
the one loop corrections in such a realistic case carefully, one gets no late time divergences \cite{Sen-Zar} and the loop 
integral receives most contributions from times near the time at which the mode is crossing the Hubble radius. 
So, by putting a cut-off in time integral (a little after Hubble crossing) in the toy field theory with quartic interactions 
we are attempting to simulate this.

In renormalized pertrubation theory, the bare Lagrangian (consisting of bare field, bare mass, bare self coupling and bare (non-minimal)
coupling to gravity ($\xi R \sigma^2$ ) all of which depend in the UV regulator) is to be partitioned into a renormalized Lagrangian 
and a counter-term Lagrangian, the way this partition is done determines the renormalization scheme choice. 
For scalar fields, in four dimensions, at one-loop (and polynomial interactions), there is no field 
strength renormalization. 
Since we wish to keep the field massless and minimally coupled, the physical mass and physical value of $\xi$ should both vanish.
Unless we are using the so called On-Shell (OS) renormalization scheme, the value of the parameters in the renormalized Lagrangian 
need not be the physical values of those parameters. One can implement a form of minimal subtraction renormalization scheme in a 
fairly straighforward way.
One can easily regularize the UV divergences by putting a cut-off in three 
dimensional momentum space (and tune the corrections such that the field stays massless). 

\begin{figure}[t]
{\includegraphics[width=8cm]{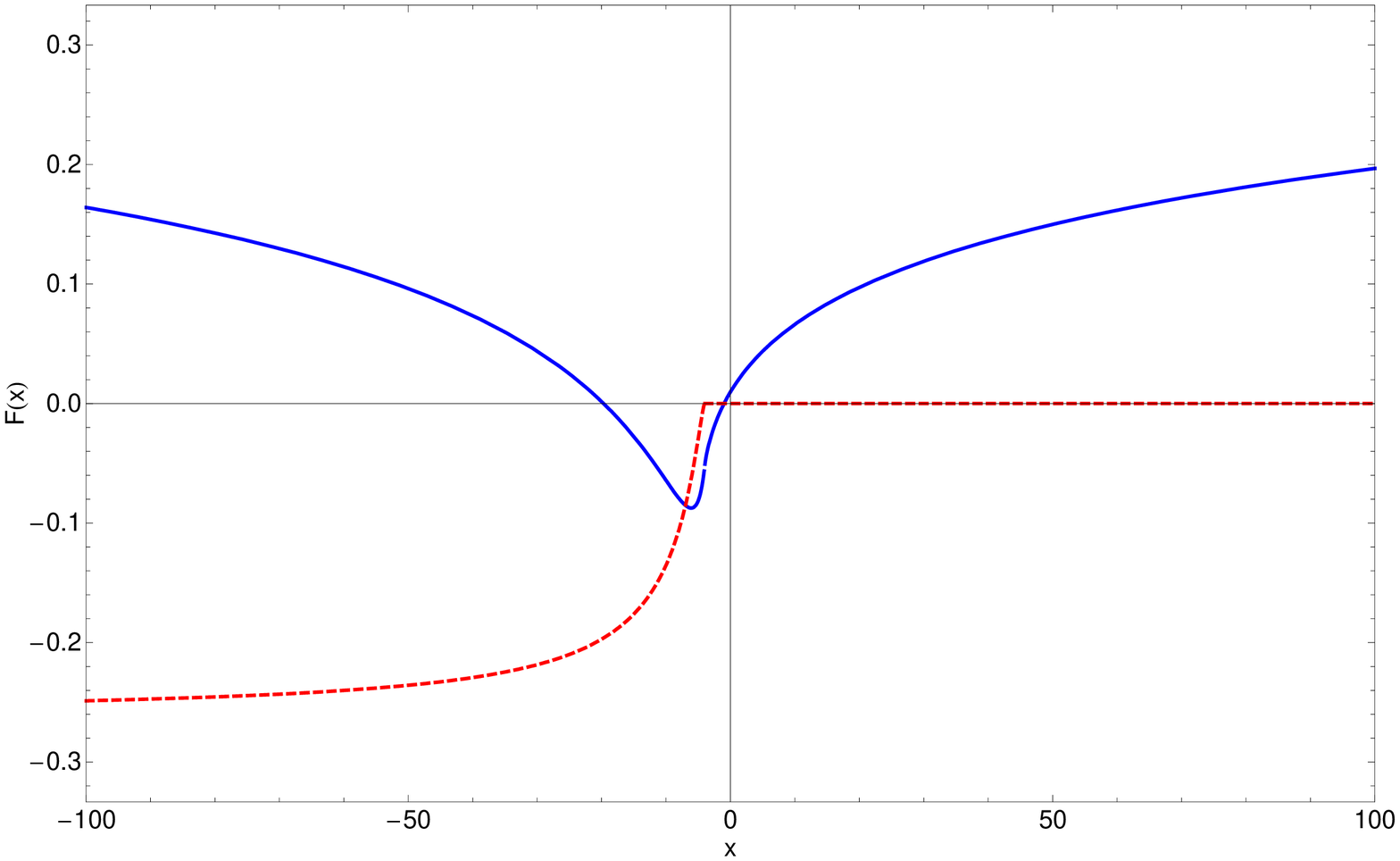}}
{\includegraphics[width=8cm]{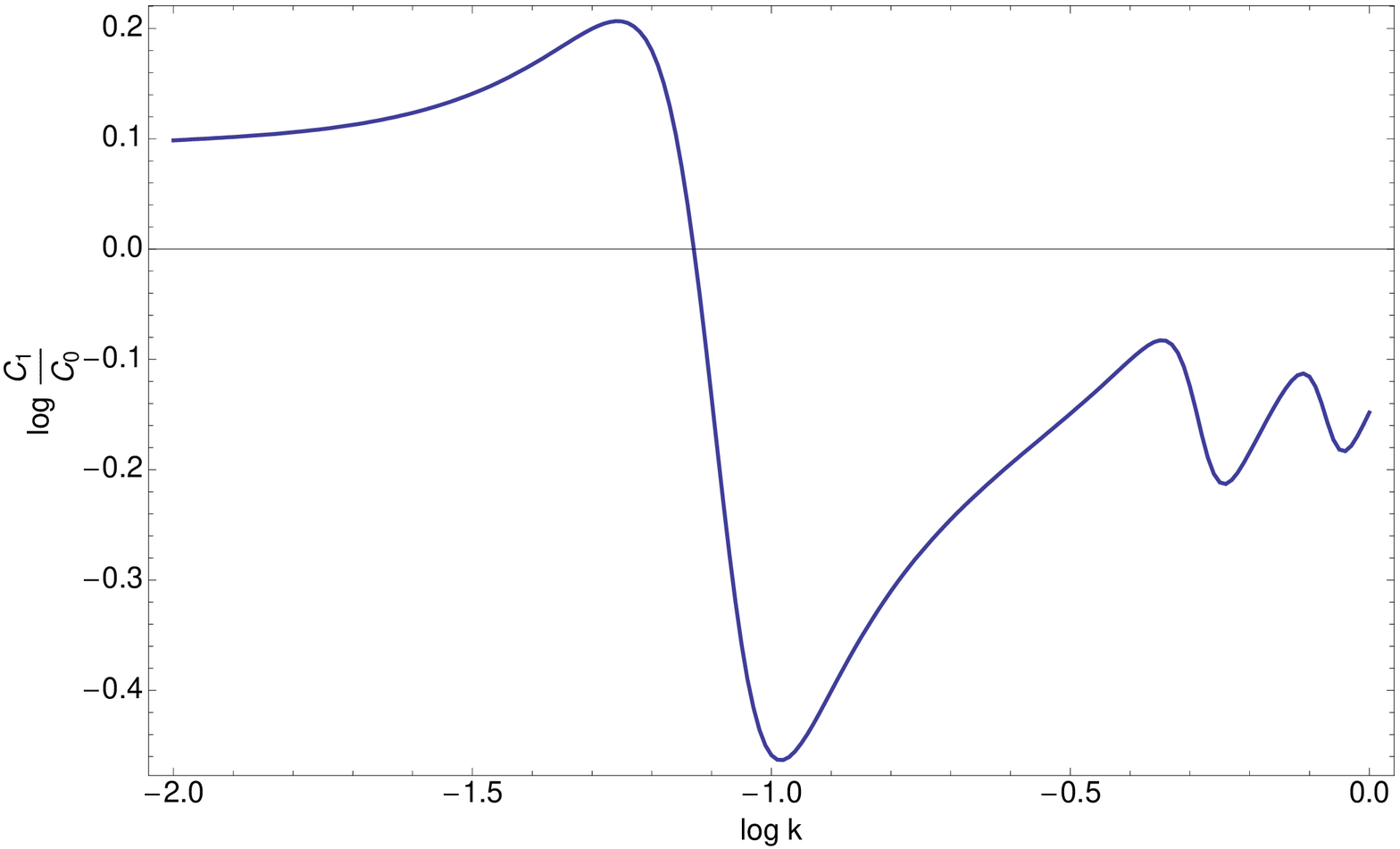}}
\caption{\small\sf \emph{Left:} 
For $\phi^3$ theory in $6$ spacetime (flat) dimensions:
The variation of real part of $F(x)$ (blue, continuous) and imaginary part of $F(x)$ (red, dashed) 
against $x = k^2/m^2$.
This function has modest values even when $x$ becomes large. 
\emph{Right:} For $\sigma$ theory (Eq. (\ref{sigaction})):
Logarithm of the ratio $C_1/C_0$ as a function of $\log k$ for the case in which $A_-/A_+$ is $0.2$ in MS 
renormalization scheme. Notice that $C_1$ increases with respect to $C_0$ and takes up its largest value at a scale 
just larger than the scale at which the dip in the tree-level power spectrum arises. For this value of $A_-/A_+$, the enhancement is 
very small, $C_1/C_0$ increases from (roughly) $26$ percent to $58$ percent. At other values of $A_-/A_+$, this enhancement will be 
larger.}\label{F6}
\end{figure}

All this can be easily compared to what happens in the usual flat spacetime field theory. Consider $\phi^3$ theory in $d=6$ spacetime 
dimensions (the theory doesn't have a true ground state but perturbation theory doesn't ``know'' about it). 
Suppose we evaluate the one-loop corrections to the (four dimensional Fourier 
transform of) two-point function in this theory. We know the tree level answer, which is (we are using the signature (-,+,+,+) and 
$x = k^2/m^2$) $\Delta(k^2) = \frac{1}{k^2+m^2} = \frac{1}{m^2(1+x)}$. 
Loop corrections are of the order of $g^2$ and if we call $g^2/(4\pi)^3$ to be $\alpha$, then the two-point function (one-loop) is

\begin{equation}
 \Delta(x) = \frac{1}{m^2(1+x)}\left[ 1 + \alpha F(x) \right]
\end{equation}

\noindent since we are working with a weakly coupled theory, we could be working with (say) $g = 0.1$ and then 
$\alpha = {\cal O} (10^{-8})$. Since the next order correction is supposed to be of the order of $10^{-16}$, even if $F(x)$ 
becomes fairly large, perturbative calculation shall remain valid. Now we use dimensional regularization and 
OS renormalization scheme to get the function $F(x)$ as shown in the left part of Fig. (\ref{F6}). 
What we are calculating is the equivalent of this for the $\sigma$ field. 
There exist a small range of modes, near the dip in the tree-level power spectrum, where the tree-level contribution is at its 
lowest value while the one-loop contribution is not so small, this causes a relative enhancement in the value of $C_1$ w.r.t 
$C_0$ (see Fig. (\ref{F6})). The result is shown for the case in which the value of the parameter $A_-/A_+$ of \cite{Staro1992} 
is $0.2$. As one changes this parameter, the depth of the dip in tree level spectrum can be increased thus increasing the relative 
size of loop corrections. We avoid considering cases in which the depth in the dip is too deep because in those cases, the loop 
correction shall become very large signalling break down of perturbation theory. 
In conclusion, we attempted to study how the 
loop corrections to the correlation functions of cosmological perturbations get affected
by the actual dynamics of the mode functions. We chose the mode functions of the field such that the tree level power spectrum has a sharp
dip and we realized that the power spectrum at one loop, renormalized in MS scheme can get enhanced (at scale near the tree level dip) 
relative to the tree level power spectrum. 

\end{document}